\documentclass[twocolumn,showpacs,preprintnumbers,amsmath,amssymb]{revtex4}

\usepackage{graphicx}% Include figure files
\usepackage{dcolumn}% Align table columns on decimal point
\usepackage{bm}% bold math
\usepackage{color}

\begin{document}

\preprint{APS/123-QED}

\title{Universal Features of Quantized Thermal Conductance of Carbon Nanotubes}
\author{Takahiro Yamamoto$^{1,3}$}
%\email[e-mail: ]{takahiro@rs.kagu.tus.ac.jp}
\author{Satoshi Watanabe$^{2,3}$}
\author{Kazuyuki Watanabe$^{1,3}$}
\affiliation{
$^{1}$Department of Physics, Tokyo University of Science, 1-3 Kagurazaka, Shinjuku-ku, Tokyo 162-8601, Japan}
\affiliation{
$^{2}$Department of Material Engineering, University of Tokyo, 7-3-1 Hongo, Bunkyo-ku, Tokyo 113-8656, Japan}
\affiliation{
$^{3}$CREST, Japan Science and Technology Agency, 4-1-8 Honcho Kawaguchi, Saitama 332-0012, Japan.}

\date{
\today
}% It is always \today, today,
             %  but any date may be explicitly specified
\begin{abstract}
The universal features of quantized thermal conductance of carbon nanotubes (CNTs) are revealed through theoretical analysis based on the Landauer theory of heat transport. The phonon-derived thermal conductance of semiconducting CNTs exhibits a universal quantization in the low temperature limit, independent of the radius or atomic geometry. The temperature dependence follows a single curve given in terms of temperature scaled by the phonon energy gap. The thermal conductance of metallic CNTs has an additional contribution from electronic states, which also exhibits quantized behavior up to room temperature.
\end{abstract}

%81.07.De Nanotubes 
%44.10.+i Heat conduction (see also 66.60.+a and 66.70.+f in transport properties of condensed matter) 
%63.22.+m Phonons or vibrational states in low-dimensional structures and nanoscale materials 
 
%\keywords{Suggested keywords}%Use showkeys class option if keyword
                              %display desired
\maketitle
During the last two decades, electronic transport in mesoscopic systems whose dimensions are much smaller than a mean-free-path of electrons has been extensively studied experimentally and theoretically. One of the most striking phenomena is that the electrical conductance is quantized in multiples of universal quantum, $2e^2/h$, which is observed in a one-dimensional (1D) conductor formed between two reservoirs~\cite{rf:datt}.

In contrast, only few studies on nano-scale thermal transport phenomena, especially at low temperatures, have been performed, because the experimental observation and evaluation of thermal transport quantities under such conditions are hard to carry out.  Recent advances in nanotechnology, however, have made it possible to investigate thermal transport phenomena in mesoscopic and nano-scale ballistic systems experimentally. Schwab {\it et al.} observed the universal quantum of thermal conductance, $\pi^2k_B^2T/3h$, in nano-sized narrow wires using a sophisticated fabrication technique~\cite{rf:schw}, and the value observed is consistent with the theoretical prediction proposed earlier by Rego and Kirczenow~\cite{rf:rego} and by other theoretical studies~\cite{rf:ange,rf:blen}. However, an unequivocal verification of the existence of quantized thermal conductance in other nanostructures has not been obtained thus far, in contrast to the quantization of electronic transport, which has been observed in various systems. As local heating is one of the key issues to be resolved in the development of nano-scale devices, it is highly desired to clarify the thermal properties of nanostructures.

Single-wall carbon nanotubes (CNTs), which are natural quantum wires of small size and high stiffness, are considered most suitable for this purpose, 
having a large phonon-mean-free-path of the order of $1~\mu$m~\cite{rf:xiao,rf:hon1}.
Recent experiments showed that CNTs have remarkable properties of thermal transport at low temperatures, reflecting quantum effects in one dimension~\cite{rf:hon1,rf:hon4,rf:hon3}.
The low-temperature thermal conductance of CNTs is linear in the temperature $T$ and extrapolates to zero at $T=0$. This implies the existence of quantized thermal conductance in CNTs.

In this Letter, we investigate the low-temperature thermal conductance in single-wall CNTs sandwiched between hot and cold heat baths, and show not only that the thermal conductance in CNTs is quantized, but also that the phonon contribution to the thermal conductance has universal features that depend only on the tube radius and not on the chirality. In contrast, the thermal conductance by electrons is found to be critically dependent on the chirality, equivalent to the electronic states.

As in previous theoretical studies on quantized thermal conductance~\cite{rf:rego,rf:ange,rf:blen}, the present study begins with a method analogous to the Landauer theory of electrical transport, since a crude formulation of phonon thermal conductance based on a conventional argument of the kinetic theory of gases is not applicable to ballistic phonon systems. The thermal current density of the 1D phonon system formed between two (hot/cold) heat baths is described as the Landauer energy flux, which is given by
%==================== Equation (1) ==============================%
\begin{eqnarray}
{\dot Q}_{\rm ph}&=&\sum_m\int_{0}^{\infty}\frac{dk}{2\pi}
\hbar\omega_m(k)v_m(k)\nonumber\\
& &\times[\eta(\omega_m,T_{\rm hot})-
\eta(\omega_m,T_{\rm cold})]\zeta_m(k),
\label{eq:1}
\end{eqnarray}
%=================================================================%
where $m$ is a phonon mode, $\hbar\omega_m(k)$ is a phonon energy dispersion of wavenumber $k$, $v_m(k)=d\omega_m(k)/dk$ is a phonon group velocity, $\eta(\omega_m,T_\alpha)=\left[\exp\left(\hbar\omega_m/k_BT_\alpha\right)-1\right]^{-1}$ is the Bose-Einstein distribution function of phonons in heat baths, and $\zeta_m(k)$ is the transmission probability between the system and heat baths~\cite{rf:rego}.

%%%%%%%%%%%%%%%%%%%%%%%%%% Fig.1 %%%%%%%%%%%%%%%%%%%%%%%%
\begin{figure}[t]
  \begin{center}
    \includegraphics[keepaspectratio=true,height=60mm]{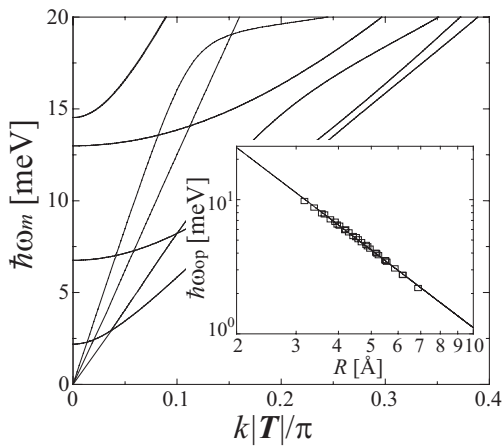}
  \end{center}
  \caption{Low-energy phonon dispersion curves for a $(10,10)$ carbon nanotube. 
  Four acoustic modes are present: a longitudinal acoustic mode, doubly degenerate 
  transverse acoustic modes, and a twisting acoustic mode (in order from the top). 
  The inset shows the energy gap $\hbar\omega_{\rm op}$ of the lowest 
  optical modes.
  }
  \label{fig:1}
\end{figure}
%%%%%%%%%%%%%%%%%%%%%%%%%%%%%%%%%%%%%%%%%%%%%%%%%%%%%%%%%

The evalution of the integration in Eq.~(\ref{eq:1}) is, in general, very difficult, and it requires a knowledge of the frequency, $\omega_m(k)$, and the transmission probability, $\zeta_m(k)$, as a function of $m$ and $k$. However, changing the integration variable in Eq.~(\ref{eq:1}) from $k$ to $\omega_m$ leads to a cancellation between the phonon group velocity, $v_m$, and the density of state, $dk/d\omega_m$, and Eq.~(\ref{eq:1}) is rewritten as
%==================== Equation (2) ==============================%
\begin{eqnarray}
{\dot Q}_{\rm ph}&=&\sum_m\int_{\omega_m^{\rm min}}^{\omega_m^{\rm max}}
\frac{d\omega_m}{2\pi}\hbar\omega_m \nonumber\\
& &\times\left[\eta(\omega_m,T_{\rm hot})-
\eta(\omega_m,T_{\rm cold})\right]\zeta(\omega_m),
\label{eq:2}
\end{eqnarray}
%=================================================================%
where $\omega_m^{\rm min}$ and $\omega_m^{\rm max}$ are the minimum and maximum angular frequencies of the $m$-th phonon dispersion, respectively. It is noted that this form is independent of the energy dispersion except for $\omega_m^{\rm min}$ and $\omega_m^{\rm max}$.
Furthermore, imposing the condition of the limits of linear responce, $\Delta T\equiv T_{\rm hot}-T_{\rm cold}\ll T\equiv(T_{\rm hot}+T_{\rm cold})/2$, and the limit of adiabatic contact between the system and heat baths, $\zeta(\omega_m)=1$, the thermal conductance, $\kappa_{\rm ph}={\dot Q}_{\rm ph}/{\Delta T}$, is given as a form of an elementary integration:
%==================== Equation (3) ==============================%
\begin{eqnarray}
\kappa_{\rm ph}=
\frac{k_B^2T}{2\pi\hbar}\sum_m\int_{x_m^{\rm min}}^{x_m^{\rm max}}
dx\frac{x^2{\rm e}^x}{\left({\rm e}^x-1\right)^2}.
\label{eq:3}
\end{eqnarray}
%=================================================================%
Performing the integration in Eq.~(\ref{eq:3}), we finally derive an analytical form of the thermal conductance, which is easily applicable to various 1D ballistic phonon system, $\kappa_{\rm ph}=\kappa_{\rm ph}^{\rm min}-\kappa_{\rm ph}^{\rm max}$:
%==================== Equation (4) ==============================%
\begin{eqnarray}
\kappa_{\rm ph}^{\alpha}&=&\frac{2k_B^2T}{h}\sum_m
\left[\phi(2,e^{-x_m^\alpha})\right.\nonumber\\
& &\left.+x_m^\alpha\phi(1,e^{-x_m^\alpha})
+\frac{(x_m^\alpha)^2}{2}\eta(x_m^\alpha)
\right].
\label{eq:4}
\end{eqnarray}
%=================================================================%
Here, $\alpha$ denotes `min' or `max', $\phi(z,s)=\sum_{n=1}^{\infty}(s^n/n^z)$ is the Appel function and $x_m^\alpha=\hbar\omega_m^\alpha/k_BT$. In particular, a gapless mode with $\omega_m^{\rm min}=0$ contributes a universal quantum of $\kappa_0=\pi^2k_B^2T/3h$ to the thermal conductance.

The thermal conductance in CNTs can be obtained by substituting $\omega_m^{\rm min}$ and $\omega_m^{\rm max}$ for all phonon modes $m$ into Eq.~(\ref{eq:4}). The phonon energy dispersions for CNTs can be obtained by diagonalizing the dynamical matrix constructed with the scaled force-constant parameters~\cite{rf:sai1,rf:sai2}. Figure~\ref{fig:1} shows energy dispersion curves for the region near the $\Gamma$ point ($k=0$) for a CNT with chiral vector ${\mbox{\boldmath $C$}}_h=(10,10)$, where $|{\mbox{\boldmath $T$}}|$ denotes the magnitude of the unit vector along the tube axis. Here, the chiral vector $(n,m)$ uniquely determines the geometrical structure of CNTs~\cite{rf:sai1,rf:dres}. The figure shows four acoustic modes with linear dispersion: a longitudinal acoustic mode, doubly degenerate 
transverse acoustic modes, and a twisting mode. The lowest optical ($E_{2g}$ Raman active) modes are doubly degenerate and have an energy gap of $\hbar\omega_{\rm op}=2.1$ meV at the $\Gamma$ point. The gap, $\hbar\omega_{\rm op}$, depends only on the tube radius $R$ and decreases approximately according to $\sim 1/R^2$ as shown in the inset of Fig.~\ref{fig:1}~\cite{rf:sai1,rf:sai2}. These modes always lie in low-energy dispersion relations, independent of the tube geometry such as radius or chirality.

%%%%%%%%%%%%%%%%%%%% Fig.2(a) and (b) %%%%%%%%%%%%%%%%%%%
\begin{figure}[t]
  \begin{center}
    \includegraphics[keepaspectratio=true,height=120mm]{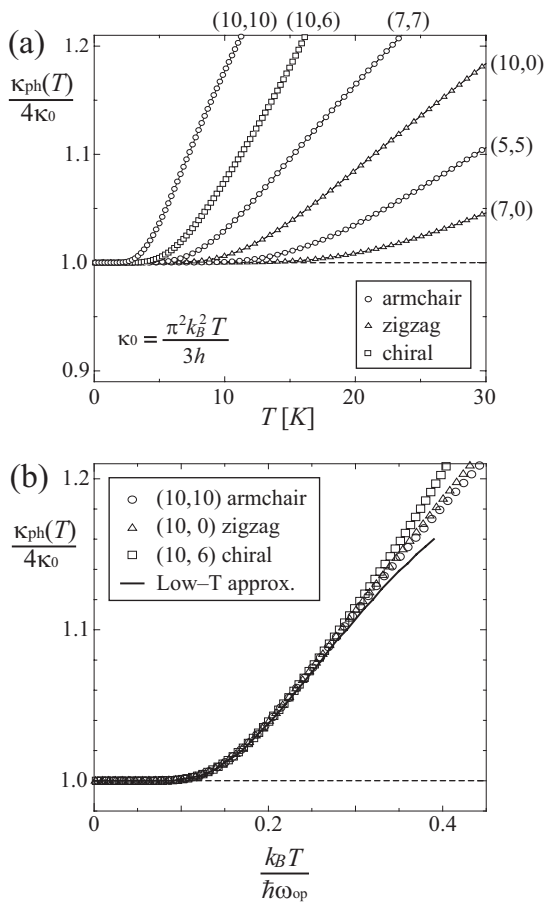}
  \end{center}
  \caption{(a) Low-temperature phonon-derived thermal-conductance for various types of 
  carbon nanotubes (CNTs) with several chiral vectors $(n,m)$. (b) Thermal conductance
  as a function of temperature scaled by the energy gap of the lowest optical mode.}
  \label{fig:2}
\end{figure}
%%%%%%%%%%%%%%%%%%%%%%%%%%%%%%%%%%%%%%%%%%%%%%%%%%%%%%%%

Figure~\ref{fig:2}(a) shows the temperature-dependent thermal conductances, normalized to a universal value of $4\kappa_0$ (as explained later). The calculated values approach unity in the low-temperature limit, indicating that the phonon thermal conductance of CNTs is quantized according to a universal value of $4\kappa_0$, independent of the shapes or atomic geometry. The quantization of thermal conductance originates from low-energy excitations of long-wavelength acoustic phonons (four branches in Fig.~\ref{fig:1}) at temperatures sufficiently low that the two lowest optical modes with an energy gap, $\hbar\omega_{\rm op}$, are not excited (lowest gapped branch in Fig.~\ref{fig:2}). The quantization can also be derived analytically from Eq.~(\ref{eq:4}). Only the first term, $\phi(2,1)$, contributes to the conductance at the low-temperature limit, leading to $4(\pi^2k_B^2T/3h)=4\kappa_0$. Here, $4$ represents the number of acoustic branches.

It may be difficult to observe the above features experimentally, because of the effects of phonon reflections at the contacts where the CNT connects to the heat baths. We, however, expect that the observation of the above feature must be possible, judging from the theoretical study of Rego and Kirczenow~\cite{rf:rego} which shows that, for quantum wire with small cross-section, reflectionless contact is realized when the connection is extremely smooth. For CNTs, such a smooth contact on the atomic scale may be formed between a silicon carbide (SiC) surface and a CNT grown on it, because the interface between the SiC and CNT may have little strain owing to their similar bond lengths. Of course, detailed theoretical and numerical works are necessary in order to fully understand the effects of phonon reflections at the contacts on thermal conductance in CNTs

Another important finding is that the different curves of temperature-dependent thermal conductance for various CNTs seen in Fig.~\ref{fig:2}(a) exhibit a universal behavior when a scaled temperature is introduced, $\tau_{\rm op}=k_BT/\hbar\omega_{\rm op}$. Taking account of the four acoustic and two lowest optical modes and substituting the values of $\omega_m^{\rm min}$ for these branches at the $\Gamma$ point into the formula of Eq.~(\ref{eq:2}), the thermal conductance can be given in the following simple form. 
%==================== Equation (3) ==============================%
\begin{equation}
\frac{\kappa_{\rm ph}}{4\kappa_0}\approx 1+\frac{3}{\pi^2}e^{-1/\tau_{\rm op}}
\left(1+\frac{1}{\tau_{\rm op}}+\frac{1}{2\tau_{\rm op}^2}\right).
\label{eq:5}
\end{equation}
%=================================================================%
The curves in Fig.~\ref{fig:2}(a) are replotted against the curve of Eq.~(\ref{eq:5}) with scaled temperature $\tau_{\rm op}$ in Fig.~2(b). It is evident that all curves (only three curves are shown for clarity) fall on a single curve coinciding with the curve of Eq.~(\ref{eq:5}) in the low-temperature limit. The curves turn upward at around $\tau_{\rm op}\approx 0.14$ from a linear region in this plot (quantization plateau), with the the plateau width determined by the tube radius according to the relation $\sim 1/R^2$ (see result in the inset of Fig.~\ref{fig:1}). This universal behavior of the thermal conductance of CNTs indicate that low-temperature phonon transport is characterized by the optical phonon energy gap, $\hbar\omega_{\rm op}$, which is determined only by the tube radius, as shown in the inset of Fig.~\ref{fig:1}. This theoretical result supports both the experimental observations and the inferred tube-radius dependence of the width of the thermal conductance plateau, although direct comparison of the absolute values between the experiment and theory is not possible due to the unknown extrinsic factors in the experiment~\cite{rf:hon3,rf:hon4}.

The electronic contribution to thermal conductance can be determined in a straightforward manner by replacing the Bose-Einstein distribution function, $\eta(\omega_m)$, in Eq.~(\ref{eq:1}) with the Fermi-Dirac distribution function, $f(\epsilon_m,T)=1/\left(e^{(\epsilon_m-\mu)/k_BT}+1\right)$, and then substituting the electron energy bands, $\epsilon_m$, into the formula. According to this formulation, all conduction bands crossing the Fermi energy level yield the same universal thermal conductance ($\kappa_0=\pi^2k_B^2T/3h$) as that of phonons, even though electrons obey different statistics. In general, the quantum of thermal conductance should be a universal value independent of particle statistics~\cite{rf:rego,rf:rego2}.

%%%%%%%%%%%%%%%%%%%%%%%%%% Fig.3 %%%%%%%%%%%%%%%%%%%%%%%%
\begin{figure}[t]
  \begin{center}
    \includegraphics[keepaspectratio=true,height=60mm]{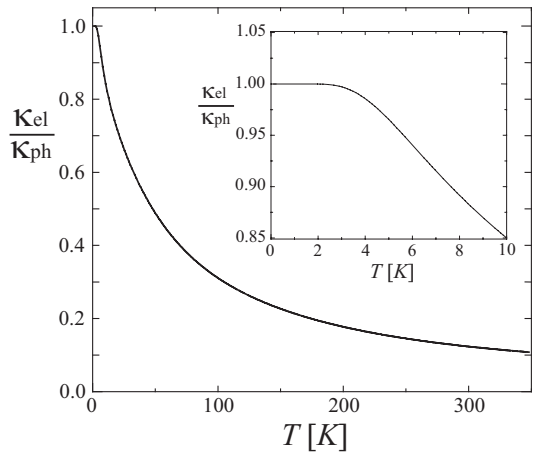}
  \end{center}
  \caption{Ratio of thermal conductance by electrons, $\kappa_{\rm el}$, 
  to that by phonons, $\kappa_{\rm ph}$, for a $(10,10)$ CNT. The inset 
  gives results at low temperatures on an expanded scale.}
  \label{fig:fig3.eps}
\end{figure}
%%%%%%%%%%%%%%%%%%%%%%%%%%%%%%%%%%%%%%%%%%%%%%%%%%%%%%%%

The behavior of the electronic thermal conductance in CNTs depends on whether the CNT is metallic or semiconducting, which is sensitive to radius and chirality~\cite{rf:sai3,rf:hama}. For semiconducting CNTs, the electronic thermal conductance should vanish roughly exponentially according to $T\to 0$, having an energy gap of the order of $0.1$~eV~\cite{rf:dre2,rf:olk,rf:wil}. For metallic CNTs, two linear energy bands crossing the Fermi level at $k>0$ \cite{rf:sai1} contribute to the electronic thermal conductance at low temperatures, resulting in a universal value of $\kappa_{\rm el}=4\kappa_0$, where the factor $4$ represents the number of two spin-degenerate channels crossing the Fermi level. This result also satisfies the Wiedemann-Franz relation between electrical conductance and electronic thermal conductance~\cite{rf:gutt,rf:grei,rf:mole}. The total thermal conductance of metallic CNTs is given by $\kappa=\kappa_{\rm el}+\kappa_{\rm ph}=8\pi^2k_B^2T/3h$ at low temperatures.

Finally, a significant difference was recognized between the widths of the quantization plateau for phonons and electrons in metallic CNTs. The characteristic energy for phonon transport at low temperature is $\hbar\omega_{\rm op}$, typically a few meV, as described in Fig. 2(b). However, the characteristic energy for electrons is of the order of $0.1$~eV, corresponding to the energy at a van Hove singularity measured from the Fermi level~\cite{rf:reo}. Consequently, the quantized nature of thermal conductance caused by electrons is predicted to survive up to room temperature, at which phonons already cease to exhibit thermal quantization, giving rise to high thermal conductance. In other words, the electronic contribution to thermal conductance is negligible compared to that from phonons at moderate temperatures. Figure~3 illustrates the temperature dependence of the ratio of thermal conductance $\kappa_{\rm el}/\kappa_{\rm ph}$ for electrons and phonons. The ratio observed experimentally~\cite{rf:hon1} is one order of magnitude lower than the present value. The discrepancy is attributed to the theoretical treatment of CNTs as purely metallic, whereas only a certain fraction ($\sim 1/3$)~\cite{rf:sai1,rf:sai3} of the crystalline ropes of CNTs in the experiment will be metallic and contribute to $\kappa_{\rm el}$.

It is noted that decisively different behaviors of thermal conductance are seen between single-wall and multi-wall CNTs. In the case of multi-wall CNTs, recent measurements showed that their thermal conductivity increases as $\sim T^{2}$ at low temperature~\cite{rf:yi,rf:kim}, suggesting that they act essentially as two-dimensional systems for phonon transport, and displays a single peak near the room temperature due to Umklapp scattering~\cite{rf:kim}. On the other hand, such a peak has not been observed so far for single-wall CNTs. This, together with the large phonon-mean-free-path of the order of $1\mu$m~\cite{rf:xiao,rf:hon1} for single-wall CNTs, justifies the neglect of the effect of the Umklapp scattering process in our approach. To develop a theory for multi-wall CNTs with taking the process into account will be a challenge for future studies.

In summary, the universal features of quantized thermal conductance in single-wall CNTs have been shown using a simple formula derived from the Landauer formalism of heat transport. Essentially, the gap of the lowest optical mode depending only on the tube radius characterizes the universal behavior of low-temperature thermal conductance by phonons in CNTs. Electrons similarly contribute a universal quantum to the thermal conductance in metallic CNTs.
The formula derived in the present study will facilitate study on the origin of thermal conductance for various one-dimensional nanostructures at low temperatures through consideration of the universal features of thermal conductance as a function of temperature scaled by the energy gaps of electrons or phonons at the long-wavelength limit. This line of study is expected to afford guidelines for the thermal management of various nano-scale electronic devices, including CNTs, which will be crucial in the assembly of fully functional devices in the future.

The present study was partly supported by the Ministry of Education, Sports, Culture, Science and Technology of Japan through the Grant-in-Aid No.~15607018.

%\end{document}

\end{document}